# Predicting Performance of Microfluidic-Based Alginate Microfibers with Feature-Supplemented Deep Neural Networks

Nicholus R. Clinkinbeard[1*], Justin Sehlin[1], Meharpal Singh Bhatti[1], Marilyn McNamarra[1], Reza Montazami[1], and Nicole N. Hashemi[1*]

[1] Department of Mechanical Engineering, Iowa State University, Ames, IA, 50011, USA

* Corresponding authors: nastaran@iastate.edu, Nicholus.Clinkinbeard@collins.com

**Abstract**

*Selection of solution concentrations and flow rates for the fabrication of microfibers using a microfluidic device is a largely empirical endeavor of trial-and-error, largely due to the difficulty of modeling such a multiphysics process. Machine learning, including deep neural networks, provides the potential for allowing the determination of flow rates and solution characteristics by using past fabrication data to train and validate a model. Unfortunately, microfluidics suffers from low amounts of data, which can lead to inaccuracies and overtraining. To reduce the errors inherent with developing predictive and design models using a deep neural network, two approaches are investigated: dataset expansion using the statistical properties of available samples and model enhancement through introduction of physics-related parameters, specifically dimensionless numbers such as the Reynolds, capillary, Weber, and Peclet numbers. Results show that introduction of these parameters provides enhanced predictive capability leading to increased accuracy, while no such improvements are yet observed for design parameter selection.*

1. **Introduction**

Over the past several years, evolving technology enabling fabrication of microfibers via microfluidics has drastically improved the ability of researchers to deposit cells to stable biocompatible structures for the purpose of studying such tissue-related phenomena as brain cell behavior and assessment of drug delivery systems [1]-[8]. Physical characteristics of particular importance to the success of these microfibers include their fabrication as solid or hollow, cross-sectional geometry, porosity for controlling distribution of cells within the structure and transport of nutrients, and mechanical properties that drive elasticity and strength [9]-[13]. For microfibers generated using a microfluidic chip, these characteristics have been shown to be directly related to prepolymer (cladding), core, and sheath solution composition and viscosity, fluid input flow rates, mode of polymerization, and bath solution composition. However, owing to the principally experimental state of microfiber fabrication using microfluidics, for these studies discovery of manufacturing parameter values leading to desired fiber characteristics has been dominated by a trial-and-error methodology. Much of this is due to the multiphysics nature of fiber manufacture, which relies on microfluid dynamics to provide focusing of solutions coupled with chemical reactions that result in polymerization, with the end result being a solid microfiber [3], [14]. This empirically-driven process therefore does not yet easily lend itself to predictive modeling techniques. As such, convergence on a desired set of inputs that yield beneficial fiber characteristics requires data. Because microfiber generation through microfluidics is currently in a highly experimental state, machine learning (ML) approaches become attractive for discovering relationships and building predictive models in the absence of a unified theory-based simulation approach. Unfortunately, however, available data is sparse. This limits and even hinders the usefulness of ML techniques—such as neural networks (NN)—for establishing relationships among variables [15]-[18]. Apart from sheer accuracy, one of the chief issues is poor generalization, that is, inability of a DNN-based model to make accurate predictions using parameters outside the range of values for those used to train the network. One way to increase the amount of data is to synthesize datapoints based on existing information [20]-[25]. Although pervasive in image processing, this was found to be effective for predicting performance of solid alginate microfibers [20]. While not ideal—especially when faced with only a handful of measurements—this provides a basis that can be built upon as more and more data become available.

Another concept used to improve the predictive power of neural-network-based models is to develop such models using data enhanced with physics and physical-based features. Often termed physics-guided machine learning (PGML), physics-informed machine learning (PIML), or some other similar descriptive [21], the addition of physics to a machine learning model can provide a constraint on results. Pawar et al. introduced an approach by using simplified theories to enhance models generated by neural networks [18]. Clinkinbeard and Hashemi [19] further examined this method and discovered that not only does the generation of meaningful supplemental data based on simplified theories provide improved results over baseline neural networks, it outperforms other regressive techniques, such as linear regression, cubic support vector machine (SVM), and a trilayered neural network.

Simplified relationships in the form of dimensionless numbers developed using Buckingham's Pi theorem can provide a way to inform a neural network while not taking the rigor of enforcing natural physical laws within its framework [26]. Of particular significance for microfluid flow are the Reynolds, Peclet, and Capillary numbers [28].

The primary objective of this study is to develop a methodology whereby important features of hollow microfibers generated with a microfluidic device are accurately predicted given relevant manufacturing parameters, such as solution flow rate ratio and prepolymer solution concentration. A corollary objective is the reverse of this first goal: facilitate the smart selection of manufacturing parameters to achieve desired microfiber features, such as porosity, cross-sectional geometry, and microstructure alignment. To achieve these objectives, the following DNN-based models are designed:

- predictive models that use flow rate ratios and solution fluid properties to estimate fiber performance, specifically porosity and cross-sectional dimensions;
- design model that use fiber porosity and cross-sectional dimensions to select fluid property values and flow rate ratios; and
- the introduction of physically-relative parameters—such as the Reynolds, capillary, Weber, and Peclet numbers—into the model generation process to improve accuracy.

The implication of realizing these objectives is a decreased microfiber time to manufacture through a reduction of the amount of experimentation required to discover advantageous fabrication inputs. As will be demonstrated through this study, improved predictive accuracy can be achieved, while improved parameter selection to give predetermined fiber performance characteristics still requires development.

## 2. Methods

This section discusses the background work leading to development of two models, one that predicts alginate microfiber features given a set of manufacturing parameters (termed herein as the *predictive model*) and one that specifies manufacturing parameters that will result in prescribed desired fiber characteristics (designated the *design model*). To lay the foundation, we first present the process of microfiber fabrication via a microfluidic device. We then define and demonstrate extraction of relevant fiber characteristics—cross-sectional geometry, porosity, and microstructure alignment—that potentially play an important role in the differentiation and use of such microfibers. After next introducing the machine learning architecture developed for processing input and output parameters—an artificial deep neural network (DNN)—we subsequently examine enhancements via integration of input characteristics based in fluid dynamics, namely, dimensionless numbers and shear/prepolymer flow and viscosity ratios. Finally, we offer model performance parameters that will be used in the next section to assess the effectiveness of the DNN-based models.

At a macrolevel, the basic flow of model generation we are trying to validate is as follows: (1) collect experimental data, (2) expand the dataset the dataset from the handful of available values to a larger sampling more suited to training a machine learning model, (3) apply the expanded dataset to a deep neural network containing physical relationships to ground the results in known physics, and (4) acquire a model that either (4a) predicts fiber performance based on input parameters or (4b) allows the design of manufacturing parameters to obtained predetermined desired performance characteristics. This overarching process is illustrated in Fig. 1.

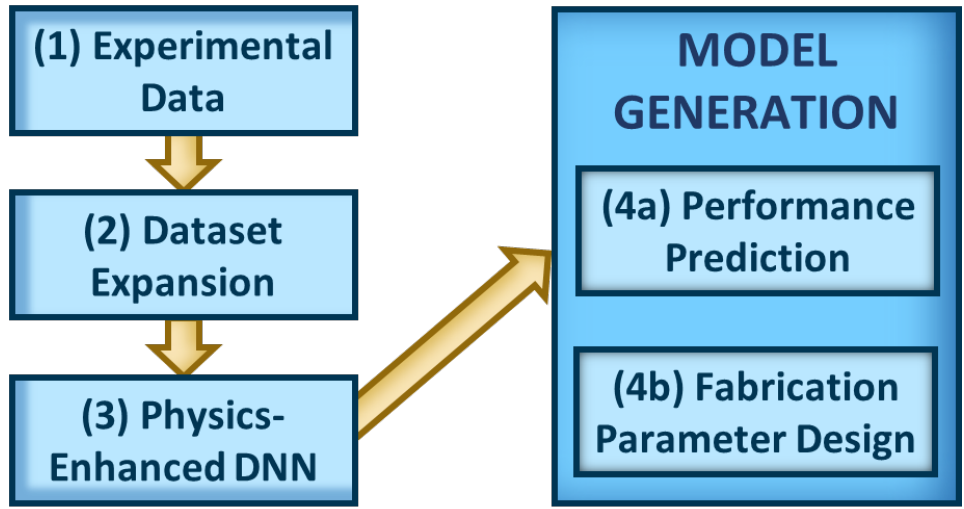

**Fig. 1.** Basic flow chart showing a desired microfluidic-based microfiber modeling process. (1) Obtain experimental data. (2) Expand experimental data by adding values that conform to the statistical properties of the original. (3) Develop a deep neural network that integrates relationships rooted in physics for enhanced accuracy. (4) Develop either a (4a) predictive or (4b) design model in order to obtain desired fiber manufacturing characteristics.

## 2.1. Experimental Foundation

Hollow alginate microfibers are fabricated for this study using a microfluidic approach, and data gathered for the training/validation and testing of a DNN-based predictive model are extracted for these structures. Well-defined manufacturing input parameters (flow rate ratios, fluid viscosity, etc.) and critical fiber features (cross-sectional geometry, porosity, microstructure alignment) are documented and used as input and output parameters to develop an effective model based on a deep neural network architecture. Although the manufacturing process for these fibers was previously introduced in [8] and [29]-[31], a description of the methodology is provided herein, followed by a discussion of the output features gleaned from the manufactured microfibers.

### 2.1.1. Fiber Fabrication

Microfibers are produced by injecting an alginate-based prepolymer/cladding solution into a microfluidic chip whereby core and sheath solutions are added to shape the fiber and induce ionic polymerization, as diagrammed in Fig. 2. The microfluidic device itself was fabricated from

poly(methyl methacrylate) (PMMA, Grainger, IL, US) to have a core channel rectangular cross-section of 1.00 mm (width) by 0.75 mm (height), as well as chevrons with 0.375 mm (width) by 0.250 mm (height). In order to achieve the channels with a Minitech CNC Mini Mill 3 Pro (Minitech Machinery Corporation, Norcross, GA), the chip was fabricated in two halves and matching sections bonded together thermally using a solvent to aid in the process.

Three fluid solutions are required to form a microfiber from sodium alginate: prepolymer/cladding, sheath, and core. These are briefly described as follows. Specific combinations of concentrations and flow rate ratios are presented in Table 1.

- Core. The core solution is a mixture of poly(ethylene glycol) (PEG, Aldrich Chemistry, St. Louis, MO, USA) and, in come cases, gelatin (Sigma Aldrich, St. Louis, MO, USA) with deionized water.

- Prepolymer/Cladding Solution. The prepolymer solution used as the basis of each manufactured microfiber is generated as a mixture of very low viscosity sodium alginate powder (Alfa Aesar, Ward Hill, MA) with deionized water. Three solution concentrations are considered: 2%, 2.5%, and 3.5% (mass/volume).

- Sheath. The sheath solution is created by mixing PEG and $CaCl_2 \cdot 2H_2O$ (Fisher Chemical, Waltham, MA, USA) with deionized water.

**Table 1.** Fluid flow rate ratios and solution concentrations used for the generation of data to inform development of a DNN-based predictive and design model. Data for fiber sets A through E were generated specifically for the present study, while data for fiber sets F through H were previously collected in support of [2]. Fiber data set I was generated contemporaneously with F through H but previously unpublished. Note that Flow Rate Ratio (FRR) is specified as Core:Prepolymer:Sheath.

| Fiber Set | Flow Rate Ratio | Core Concentration PEG/Gelatin | Prepolymer Alginate Concentration | Core Concentration PEG/CaCl$_2$-2H$_2$O |
|---|---|---|---|---|
| A | 200:150:300 | 30% / 0.08% | 3.5% | 30% / 0.75% |
| B | 200:100:300 | 30% / 0.08% | 3.5% | 30% / 0.75% |
| C | 600:250:300 | 30% / 0.08% | 3.5% | 30% / 0.75% |
| D | 500:200:1000 | 30% / 0.08% | 3.5% | 30% / 0.75% |
| E | 200:150:150 | 10% / 0.00% | 2.0% | 10% / 0.00% |
| F | 600:250:200 | 30% / 0.00% | 3.5% | 30% / 0.00% |
| G | 200:100:400 | 30% / 0.00% | 3.5% | 30% / 0.00% |
| H | 200:100:500 | 30% / 0.00% | 3.5% | 30% / 0.00% |
| I | 600:250:200 | 20% / 0.00% | 2.5% | 20% / 0.00% |

To fabricate a microfiber, each solution for a particular set is contained in syringes and injected into the microfluidic device at steady (but not necessarily equal) flow rates using one of three GenieTouch syringe pumps (Kent Scientific Corporation). As shown in Fig. 2, a single line of solution is injected into the top capillary (A), immediately followed by injection of the prepolymer/cladding solution (B) into two ports—one on each side of the main channel. Once the two fluids are in contact, they pass through a region within the main channel that contains the chevrons, which are used to hydrodynamically focus the core solution to enable generation of a hollow fiber. After passing this first set of chevrons, the sheath solution is injected to both aid in further shaping of the fiber and commence the process of ionic crosslinking that creates the alginate gel. Finally, the fledgling fiber exits the device into a 15% CaCl$_2$·2H$_2$O bath that further strengthens it. Once the fiber has entered the bath, it is manually extracted and wrapped around a paper frame for drying in a conventional deep freezer for a period not less than 24 hours. Upon removal of moisture, fiber sections are either subjected to water displacement-based porosity measurements, scanning electron microscopy (SEM) to determine cross-sectional geometry and image-based porosity calculations, or imaged for birefringence under magnification in an inverted microscope.

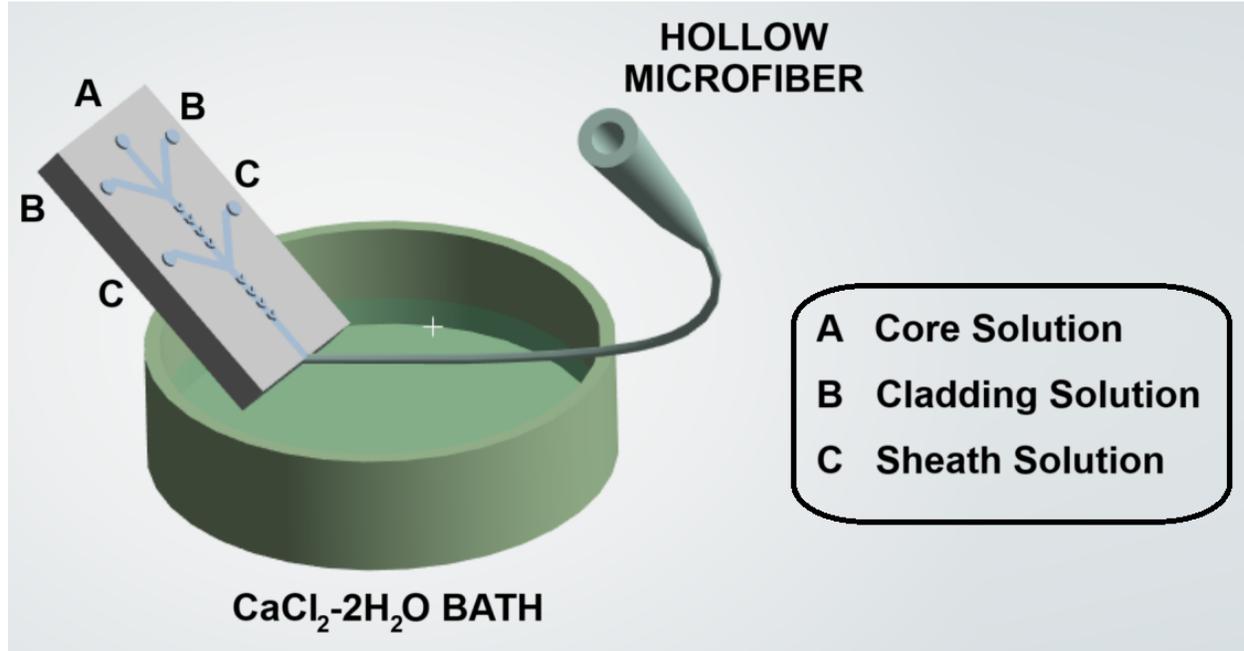

**Fig. 2.** Fabrication of an alginate-based microfiber using a microfluidic device. After preparing the prepolymer/cladding, sheath, and core solutions, these fluids are simultaneously injected into various channels of the microchip at constant (but not necessarily equivalent) flow rates using three separate syringe pumps. As the core solution (A) travels through the singular channel at the top of the device, it interfaces first with the prepolymer/cladding solution (B), which enters the chip on two opposite sides of the main channel. Immediately following the interaction of the two fluids, chevron features within the main channel attempt to hydrodynamically center the core solution to aid the creation of a hollow microfiber. After passing through the first set of chevrons, the two fluids interact with the sheath solution (C), which also helps form the fiber as ionic crosslinking begins to gel the alginate. Ionic polymerization is further aided as the newly-formed microfiber exits the device into a $CaCl_2 \cdot 2H_2O$ collection bath. The newly-formed fiber is manually retrieved from the bath and dried in a freezer for at least 24 hours prior to being subjected to porosity measurements, inspection through scanning electron microscopy, or imaged for birefringence to calculate microstructure alignment.

### 2.1.2. Solution Fluid Properties

Critical manufacturing parameters are limited not only those described in the foregoing section (i.e., solution concentrations, volumetric flow rates), but they include solution fluid property values, particularly those for alginate. Specific to this study, the following are required: density, viscosity, surface tension, and diffusion coefficient. Table 2 provides a summary of all relevant fluid property values. Density was based on measured mass and volume and calculated simply as the ratio of these two quantities. Acquisition of the remaining parameters is detailed in the following paragraphs.

**Table 2.** Fluid property values for solutions used to generate alginate microfibers. Note that all solutions comprise the solute and deionized water as the solvent and are specified in terms of solute weight per solvent mass.

| Solute Composition | Density (kg/m³) | Kinematic Viscosity (cSt) | Surface Tension (N/m) | Diffusion Coefficient ($10^{-9}$ m²/s) |
|---|---|---|---|---|
| 2.0% Alginate | 969.9 | 42.1 | 0.0286 | 0.774 |
| 2.5% Alginate | 970.0 | 95.1 | 0.0254 | 0.773 |
| 3.5% alginate | 970.0 | 200 | 0.0252 | 0.770 |
| 30% PEG / 0.08% Gelatin | 1,007 | 265 | 0.0236 | * |
| 30% PEG / 0.75% CaCl$_2$-2H$_2$O | 1,004 | 200 | 0.0277 | * |
| 10% PEG | 1,008 | 21.3 | 0.0342 | * |
| 20% PEG | 1,008 | 64.1 | 0.0283 | * |
| 30% PEG | 1,008 | 200 | 0.0226 | * |

*\*Not investigated as part of this study.*

*Viscosity.* Viscosity for each of the solutions of interest was determined using the falling sphere approach, which relates inertial, viscous, and drag forces as follows when the sphere reaches terminal velocity through the medium:

$$W = F_d + F_b \tag{1}$$

For this force balance, $W$ is the sphere weight, and $F_d$ and $F_b$ are the drag and buoyant forces on the sphere, respectively. Setting this equation in terms of its fundamental components yields

$$\rho_b V_b g = \frac{1}{2} C_D A_b \rho_s v_b^2 + \rho_s V_b g \tag{2}$$

where $\rho_b$ and $\rho_s$ are the sphere and solution densities, $V_b$ and $A_b$ are the sphere volume and cross-sectional area, $g$ is the gravitational acceleration constant, and $v_b$ is the terminal velocity of the sphere through the solution. To find terminal velocity, a small ball of known mass and diameter was dropped into the fluid and its velocity tracked through digital video. Once the ball attained terminal velocity, the drag coefficient was computed as follows:

$$C_D = \left(\frac{4}{3}\right)\left[\frac{(\rho_b - \rho_s)d_b g}{\rho_s v_b^2}\right] \tag{3}$$

where $d_b$ is the diameter of the sphere.

In order to calculate viscosity for the solution, a correlation between drag coefficient and Reynolds number of the sphere must be established. Since the connection between drag coefficient, $C_D$, and Reynolds number, $Re_S$, for the sphere falling through the fluid at terminal velocity is nonlinear, the following relationships were used [42]:

$$C_D = \frac{24}{Re_S}, \qquad Re_S < 0.2 \tag{4}$$

$$C_D = \frac{21.12}{Re_S} + \frac{6.3}{\sqrt{Re_S}}, \qquad 0.2 < Re_S < 2{,}000$$

After iteratively determining Reynolds number based on the results of Equation 4, viscosity was recovered by rearranging the Reynolds number equation [40]:

$$Re_S = \frac{\rho_b v_b d_d}{\mu_S} \Rightarrow \mu_S = \frac{\rho_b v_b d_d}{Re_S} \tag{5}$$

where $v_b$ is the terminal velocity of the sphere, $d_d$ is the sphere diameter, and $\mu_S$ is the dynamic viscosity of the solution.

*Surface Tension.* Surface tension for each solution was calculated by measurement of dimensional parameters of fluid rise through a capillary tube, as follows [42]:

$$\gamma = \frac{\rho_s g d_t h}{4 \cos \theta} \tag{6}$$

where $d_t$ is the inner diameter of the tube, $h$ is the distance the fluid into the tube above the fluid surface, and $\theta$ is the angle between the fluid and the inner capillary tube surface.

*Diffusion Coefficient.* The final parameter required is diffusion coefficient. To obtain this value for the alginate solutions, data published in [43] for an ambient temperature 298.15K was extracted and a linear curve fit for all points between a concentration (w/v) of 0.1% and 10% yielded the following relationship:

$$D = -0.2798C + 0.7796 \qquad (7)$$

where $D$ is the solution diffusion coefficient expressed as $10^{-9}$ m²/s and $C$ is the concentration expressed in decimal form.

### 2.1.3. Cross-Sectional Geometry

In order to determine cross-sectional geometry properties, samples of each batch of fibers are reserved for analysis in a scanning electron microscope (SEM), JEOL model JCM-6000. Samples are observed one at-a-time by adhering a dried portion of the fiber to the interior specimen stage via copper tape. Images of fiber cross-sections are subsequently processed using the JEOL software to measure the exterior boundary and interior hollow portion. Due to the profile of each fiber cross sections having nonconformal geometry, several measurements are made for each a given and averaged to give exterior and interior "diameters." Fig. 3 shows an example of a fiber cross section along with that for an idealized hollow microfiber to demonstrate the concepts of interior and exterior diameters.

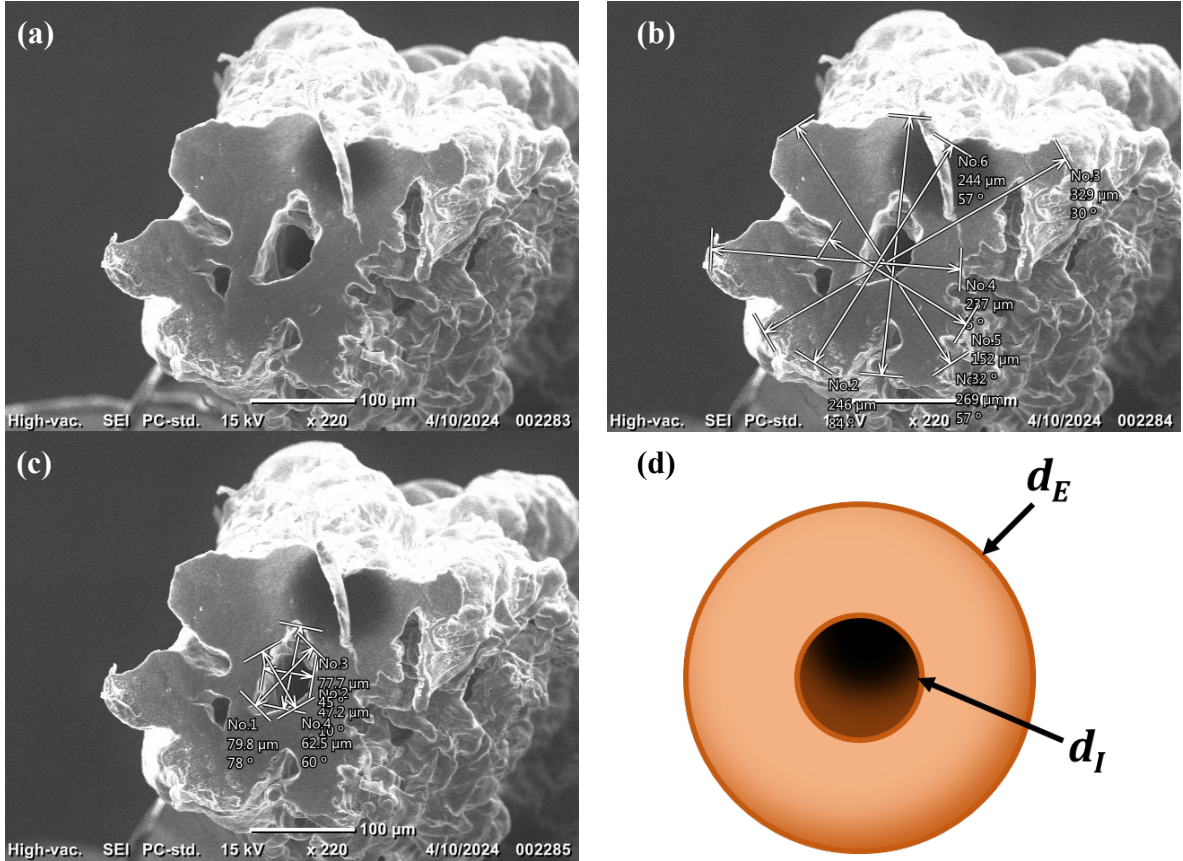

**Fig. 3.** SEM image of a hollow alginate-based microfiber cross-section (3.5% alginate solution with FRR of 200:100:300 shown). (a) Baseline image. (b) Annotated image showing various dimensions of fiber exterior boundary. (c) Annotated image showing various dimensions of fiber interior boundary. (d) Ideal hollow microfiber geometry. Note that external and internal fiber "diameters" $d_E$ and $d_I$, respectively, are estimated as the average of measured values for each geometrically nonconformal section.

A summary of microfiber mean cross-sectional dimensions is provided in Table 3.

**Table 3.** Summary of measured cross-sectional dimensions for hollow alginate-based microfibers.

| Fiber Set | FRR | Alginate Concentration | Mean Diameter (µm) Outer | Inner | Number of Samples |
|---|---|---|---|---|---|
| A | 200:150:300 | 3.5 | 191 | 53.2 | 11 |
| B | 200:100:300 | 3.5 | 212 | 43.1 | 7 |
| C | 600:250:300 | 3.5 | 306 | 102 | 10 |
| D | 500:200:1000 | 3.5 | 201 | 108 | 5 |
| E | 200:100:300 | 2.0 | 130 | 10.0 | 1 |
| F | 600:250:300 | 3.5 | 383 | 89.6 | 5 |
| G | 500:200:1000 | 3.5 | 193 | 38.0 | 3 |
| H | 200:100:300 | 3.5 | 190 | 35.6 | 4 |
| I | 200:150:150 | 2.5 | 308 | 86.5 | 5 |

### 2.1.4. Porosity

The critical parameter of fiber porosity is calculated using two methodologies. The first discussed here is a water displacement approach described by [31]. This method requires measurement of both the dry and wet mass of a particular specimen—as well as its volume—to determine a percentage value of porosity, as follows:

$$\Phi_f = \frac{m_w - m_d}{\rho V} \tag{8}$$

where $\Phi_f$ is fiber porosity, $m_d$ and $m_w$ are the dry and wet weights of the fiber, respectively, $V$ is the wetted fiber volume, and $\rho$ is the density of water used for measurements.

The process for determining porosity using this approach is to first weigh a set of dried fibers to obtain $m_d$. After subsequent immersion of the same set of fibers in deionized water, excess water (i.e., that not wholly contained by the fiber) is removed and the fiber is reweighed to determine the wet mass ($m_w$). Finally, the wet fiber is placed into a small syringe, itself containing a volume of water. The resulting volume increase within the syringe is noted as the volume of the wetted fiber.

Table 4 provides a summary of porosity values calculated using the water displacement methodology. Note that values are limited to data gathered during the course of the present study and therefore represent only a subset of data used for subsequent DNN-based predictions. As such, it is reserved for comparison with the alternative porosity calculation methodology described in the following paragraphs.

**Table 4.** Summary of hollow alginate-based microfiber porosity using the water displacement methodology.

| Fiber Set | FRR | Alginate Concentration | Mean Porosity | Std. Dev. of Porosity | Number of Samples |
|---|---|---|---|---|---|
| A | 200:150:300 | 3.5 | 21.1 | 18.8 | 12 |
| B | 200:100:300 | 3.5 | 23.6 | 13.2 | 13 |
| C | 600:250:300 | 3.5 | 30.7 | 18.8 | 13 |
| E | 200:100:300 | 2.0 | 65.2 | N/A | 1 |

The second methodology to obtaining fiber porosity is accomplished through image processing of the fiber cross-sectional SEM photographs introduced in Section 2.1.3. This approach was developed in order to determine porosity for fibers previously generated where no measurements had been taken by the water displacement method but where cross-sectional SEM photographs are available. Using an approach similar to those described by [32] and [33], each raw SEM image is segmented to remove all pixels that are not part of the cross section. This is accomplished using the LABKIT plugin that is part of the Fiji ImageJ open-source imaging package [34]. After importing the original gray image (see Fig. 4a), the region of interest is selected and LABKIT segments the image into a foreground and a background, the latter of which is masked (Fig. 4b). Upon setting a threshold for pixel darkness, all pixels in the image are converted to these binary colors to separate the porous regions (black—represented as red in Fig. 4c and Fig. 4d) from the nonporous (white—represented as blue in Fig. 4c). Finally, the porosity is calculated as the number of remaining black pixels divided by the total number of unmasked pixels.

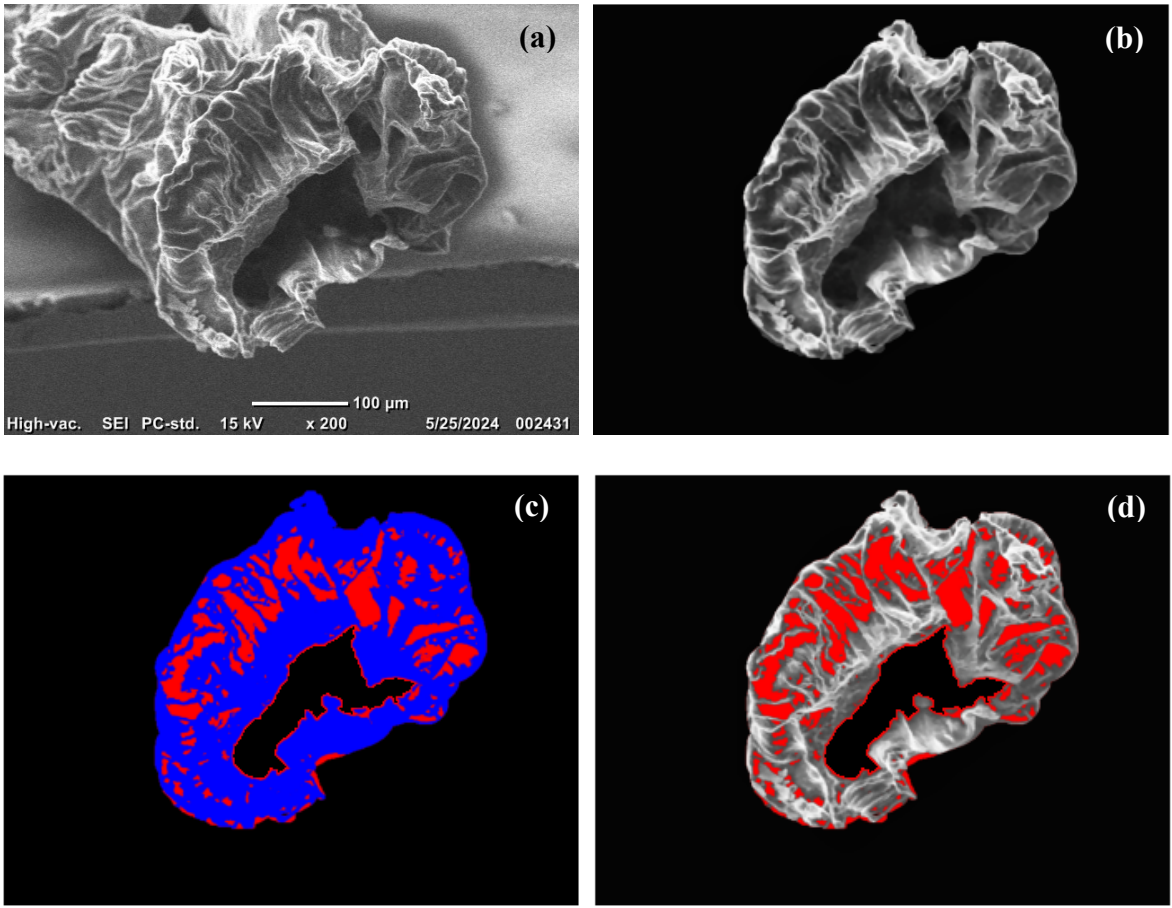

**Fig. 4.** Fiber cross-sectional images demonstrating the porosity calculation procedure using SEM images. (a) Original SEM image. (b) Segment SEM image to remove non-cross-sectional pixels. (c) Segmented image noting microfiber cross section in blue and porous regions in red. (d) Final image overlapping the porous regions with the segmented cross-sectional image. Note that the center opening was removed from SEM-based porosity calculations.

Table 5 provides a summary of porosity values calculated using the SEM image processing methodology.

**Table 5.** Summary of hollow alginate-based microfiber porosity using the SEM image processing methodology.

| Fiber Set | FRR | Alginate Concentration | Mean Porosity | Std. Dev. of Porosity | Number of Samples |
|---|---|---|---|---|---|
| A | 200:150:300 | 3.5 | 19% | 7.3% | 13 |
| B | 200:100:300 | 3.5 | 22% | 8.3% | 12 |
| C | 600:250:300 | 3.5 | 13% | 5.5% | 26 |
| D | 500:200:1000 | 3.5 | 23% | 6.4% | 5 |
| E | 200:150:150 | 2.0 | 21% | 4.6% | 8 |
| F | 600:250:200 | 3.5 | 10% | 4.0% | 5 |
| G | 200:100:400 | 3.5 | 13% | 4.1% | 4 |
| H | 200:100:500 | 3.5 | 18% | 3.0% | 4 |
| I | 600:250:200 | 2.5 | 27% | 7.4% | 14 |

In order to assess the SEM image-based porosity methodology, a comparison is made with the water-displacement approach first described. In order to do this, it must be noted that porosity completed using the water-displacement approach invariably allows for water to exist inside the hollow section, while the SEM imaging methodology corrects for this. Therefore, in order to have a direct comparison between methods, the SEM images for fiber sets A, B, and C are evaluated with the center section considered as a pore (note that set E is omitted from this assessment due to only one sample processed per Table 3). Table 6 provides a comparison of the resulting values. Note that fiber sets A and B provide very close match between the two approaches, while set C shows a 40% difference.

**Table 6.** Summary of hollow alginate-based microfiber porosity using the water displacement methodology. Note that porosity for the SEM imaging approach is adjusted to account for water inside the hollow section.

| Fiber Set | Mean Porosity, $H_2O$ Displacement | Mean Porosity, SEM Images | % Difference |
|---|---|---|---|
| A | 21.1 | 22.4 | 6.2 |
| B | 23.6 | 22.7 | 3.8 |
| C | 30.7 | 18.2 | 41 |

## 2.2. Modeling Approach

The deep neural network architecture used for this study builds on prior work [19], [20]. Although relatively straightforward to implement, a DNN is susceptible to such notable pitfalls as data sparsity. In order to address this issue, two approaches for improving the quality of data are introduced at the outset: experimental data expansion and introduction of features rooted in fluid dynamics. Once these concepts are introduced, the DNN architecture is presented. To wrap up the section, we provide DNN-based model performance evaluation criteria.

### 2.2.1. Experimental Data Expansion

Because the amount of data obtained during the experimental phase is low in comparison with what is generally necessary to develop a valid deep neural network, additional datapoints based on the statistical distributions of experimental results are synthesized, as demonstrated in Fig. 5. Since only mean and standard deviation were available for each parameter of the prior study where solid fibers were under investigation, a Gaussian distribution was assumed based on lack of better information [20]. However, because all tabular data is available for the current study with hollow microfibers, rather than simply assuming the validity of a Gaussian distribution, investigation is made to determine the best fit for synthesizing each set of data. To accomplish this, like data for a given flow rate ratio—for example, porosity with a FRR of 200:150:300—is processed through a variety of distribution functions via the Matlab command 'fitdist' in order to extract relevant statistical properties [38]. Using the function 'adtest,' the data is then subjected to the Anderson-Darling Goodnes-of-Fit Hypothesis Test [39] to assess the null hypothesis for a significance value of 5%, as well as extract the *p*-value. Those distributions with the highest *p*-value and a valid null hypothesis are selected as candidate functions for data synthesis. Rather than simply select the best fit for each independent set of data, however, effort was made to drive consistency among a particular variable by evaluating the p-values for each dataset and choosing the distribution that had the highest average overall. For example, all synthesized porosity data for all flow rates was produced to have a normal distribution, even though fiber data set I was found to have a higher p-value with a lognormal distribution. Table 7 provides the statistical distribution functions used to expand each empirical dataset.

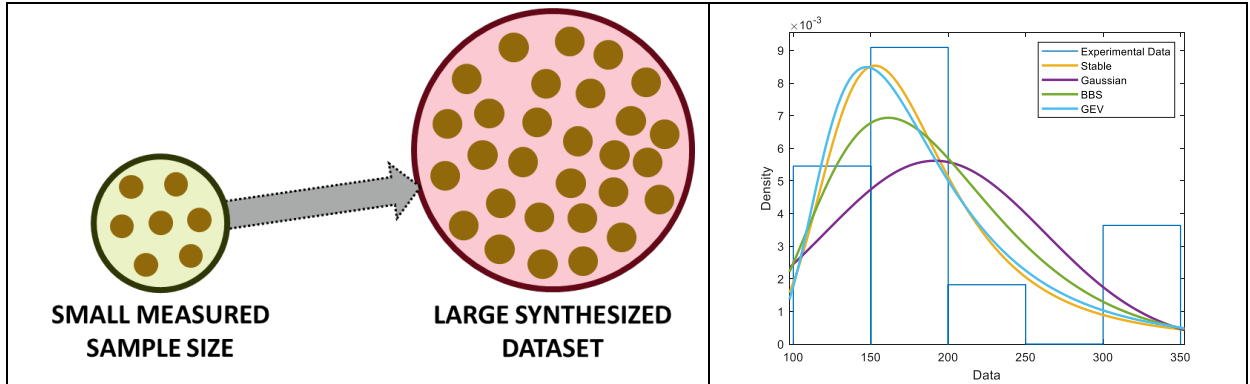

**Fig. 5.** Generation of synthetic data based on the statistical distribution and properties of empirical data collected as described in Section 2.1. (a) Schematic of mapping from sparse to large datasets. (b) Example of data fitting using different statistical distributions.

**Table 7.** Statistical distributions applied to empirical data for purpose of data expansion through synthesis.

| Fiber Feature | Distribution Function | Average p-Value |
|---|---|---|
| Outer Diameter | Gaussian | 0.785 |
| Inner Diameter | Lognormal | 0.927 |
| Porosity (SEM-Based) | Gaussian | 0.931 |
| Birefringence Order Parameter | Gaussian | 0.798 |

One notable limitation of this approach is that while when taken as a whole, each synthesized dataset conforms to its particular statistical distribution independently from values of different datasets. For example, one particular value of synthesized porosity does not correlate to particular values for, say, fiber exterior and interior diameters. However, this limitation is already inherent in the baseline experimental data upon which the synthesized is based since a single fiber section was not subjected to both porosity and cross-sectional dimension measurements. In light of this potential issue, during the course of generating expanded datasets the ratio of outer to inner diameter was calculated based on the raw data and on the expanded data and found to have similar mean values for each fiber set. Additionally, generated datasets that would be physically impossible were removed. Specifically, synthesized datasets with outer diameters less than the inner diameter were removed from the study.

### 2.2.2. Physics-Based Features

The second approach in an attempt to improve DNN-based model predictive capabilities is to introduce known physical constraints into the process rather than simply rely on a blind data-driven approach. The mechanism chosen by which physics is introduced into the DNN is the concept of dimensionless quantities based on fluid property values and flow characteristics. By using dimensionless numbers, the goal is not the end values in themselves; rather, the fact that these quantities put fluid properties and flow parameters into know relationships among one another is why they were chosen to create physical constraints on DNN-based model generation.

The following paragraphs introduce dimensionless quantities relevant to this study: Reynolds ($Re$), capillary ($Ca$), Peclet ($Pe$), and Weber ($We$) numbers.

*Reynolds Number.* The Reynolds number is the ratio of fluid inertial to viscous forces [28], with low values typically related to laminar flow while higher values can indicate turbulence. With fluid flow on the microscale—particularly with a relative viscous fluid as sodium alginate solution—laminar flow tends to dominate due to the low amount of mass that is being transport at relatively slow speeds. For the current study, however, Reynolds number is investigated as potentially enhancing the predictive power of a DNN-based model due to its function of relating various parameters, specifically fluid velocity, viscosity, and channel cross-sectional geometry. As such, it is calculated herein as

$$Re = \frac{vd_H}{\nu} \tag{9}$$

where $v$ is fluid velocity for the alginate solution upstream of the hydrodynamically focusing chevrons within the microfluidic chip, $d_H$ is the hydraulic diameter of the fluid channel, and $\nu$ is fluid kinematic viscosity. Hydraulic diameter of the channel is calculated using the dimensions provided in Section 2.1.2 for a rectangular section:

$$d_H = \frac{2wh}{w+h} \tag{10}$$

where $w$ is the channel width and $l$ is the channel height.

*Capillary Number.* The capillary number is the ratio of viscous and capillary forces within a channel and is calculated as follows [28]:

$$Ca = \frac{v\mu}{\gamma} \tag{11}$$

where $v$ is once again the alginate solution velocity through the channel, $\mu$ is dynamic viscosity, and $\gamma$ is surface tension.

*Peclet Number.* The Peclet number, *Pe*, is used to evaluate the relative significance of mass transport via diffusion as opposed to convection and is calculated as follows:

$$Pe = \frac{lv}{D} \tag{12}$$

where $l$ is a characteristic length, $v$ is the average fluid flow velocity through the channel, and $D$ is the diffusion coefficient. For purposes of this study, the hydraulic diameter used for Reynolds number is used as the characteristic length.

*Weber Number.* The final dimensionless quantity investigated for bounding DNN-based model development is the Weber number, which is calculated as follows:

$$We = \frac{\rho v^2 l}{\gamma} \tag{13}$$

where the characteristic length is once again taken as the channel hydraulic diameter.

### 2.2.3. Model Development

This section describes model development based on a deep neural network architecture.

#### 2.2.3.1. Predictive vs. Design Modeling

While the concepts are by no means unique to models generated by DNNs (or even machine learning techniques in general), because this study investigates both situations, differentiation between predictive and design models is warranted at this point. The purpose of a *predictive model* is to accept predetermined inputs and provide resultant behavior. Although such a model can take many forms, this is concept is pervasive in the use of the finite element method or computational fluid dynamics [44], [45]. For example, a structural finite element model will input boundary conditions, material properties, and forces to determine deflections, strains, and stresses. Likewise, DNN-based model of the sort studied herein used to predict microfiber performance will take manufacturing inputs (solution FRR) to estimate fiber features (cross-sectional geometry, porosity, and microstructure alignment. When used as an assessment tool, a predictive model can either provide confidence in the performance of a system design or indicate that improvements are necessary. However, as a design tool it requires many iterations with varied inputs in to allow the user to find more optimal input features.

This is where a *design model* is beneficial. For this approach, the reverse process is employed: the designer selects desired microfiber characteristics and supplies those as inputs, and the model provides appropriate manufacturing parameters to achieve these. The benefit of this methodology is to drastically reduce the amount of trial-and-error required to select appropriate manufacturing parameters.

**Fig. 6** provides conceptual diagrams of both predictive and design models.

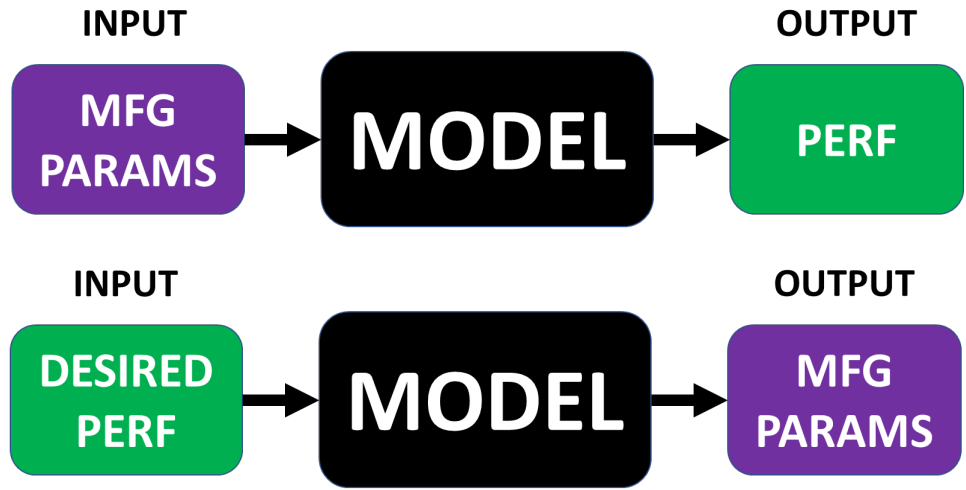

**Fig. 6.** Flow diagrams for a (a) predictive modal and (b) design model. The following abbreviations hold: "MFG" represents "manufacturing," "PARAMS" represents "parameters," and "PERF" represents "performance."

### 2.2.3.2. Deep Neural Network Architecture

Based on the predictive vs. design approach described in the previous section, six different models were generated for this study based on all nine datasets introduced in Table 1, for which fiber inner and outer diameter and porosity are available characteristics:

- <u>Case I</u>: Predictive model using fiber manufacturing solution flow rates and fluid properties as inputs. Outputs are fiber characteristics of inner and outer diameter as well as porosity.
- <u>Case II</u>: Predictive model using dimensionless numbers as inputs. Outputs are fiber inner/outer diameter and porosity.
- <u>Case III</u>: Predictive model using all solution flow rates, fluid properties, and calculated dimensionless numbers as inputs. Outputs are fiber inner and outer diameters along with porosity.
- <u>Case IV</u>: Predictive model using a DNN of the same architecture as Case III with dimensionless parameters also injected into layers 2 and 3.
- <u>Case V</u>: Design model with fiber inner/outer diameters and porosity as inputs. Outputs are dimensionless parameters.

- Case VI: Design model with fiber inner/outer diameters and porosity as inputs. Outputs are solution flow rates and fluid properties.

Fig. 7 presents the architecture for DNNs used to generate predictive models (i.e., Cases I, II, III, and IV).

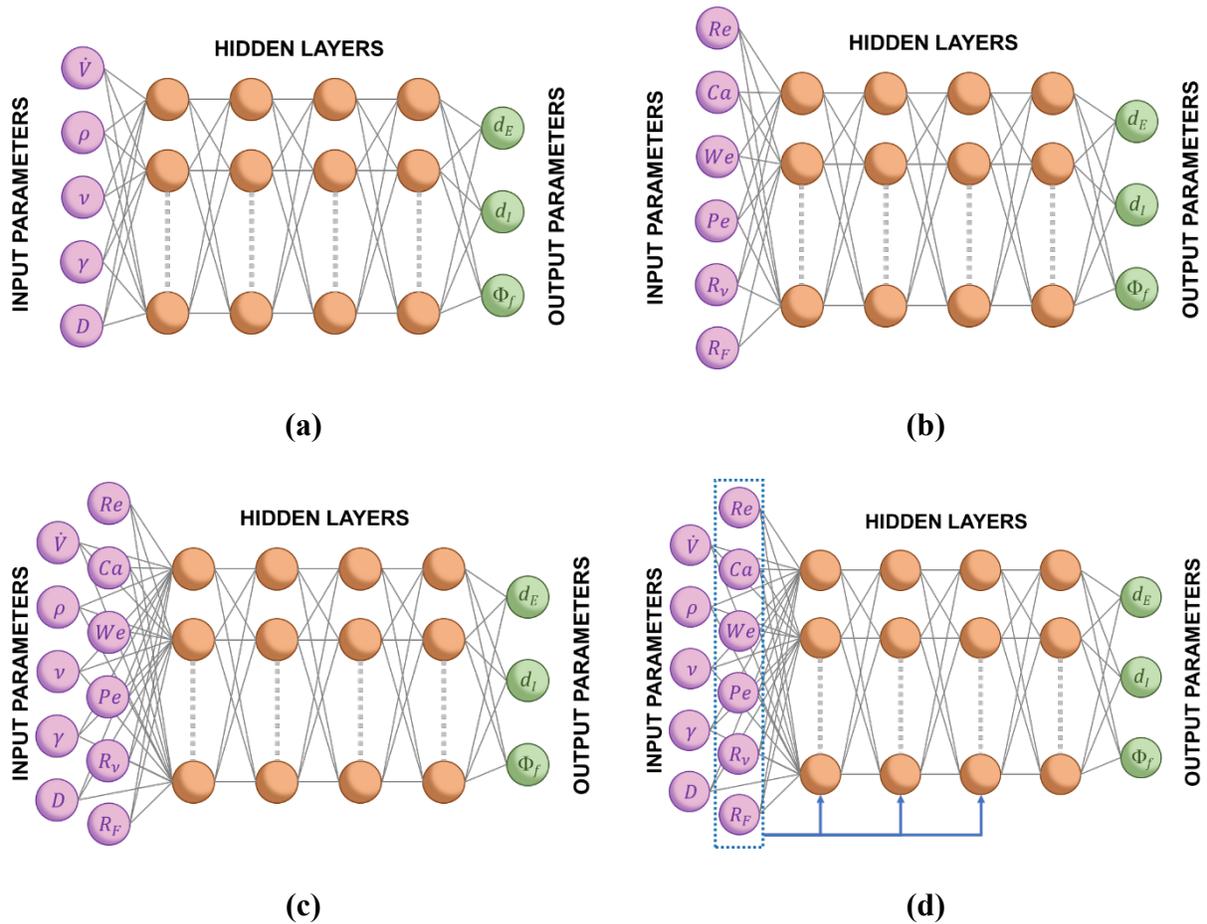

**Fig. 7.** Deep neural network architectures used to generate models for predicting fiber characteristics. (a) Case I: Generation of model based on manufacturing input parameters, i.e., core, prepolymer, and sheath flow rates, density, kinematic viscosity, surface tension, and diffusion coefficient (prepolymer only). (b) Case II: Generation of model using dimensionless numbers derived from manufacturing parameters: Reynolds number, capillary number, Weber number, and Peclet number (prepolymer only). (c) Case III: Generation of model using all manufacturing parameters and dimensionless numbers as inputs. (d) Case IV: Generation of model using all manufacturing parameters and dimensionless numbers as inputs, with the latter also injected into DNN layers 2 and 3. Note the following nomenclature: $\dot{V}$, $\rho$, $\nu$, $\gamma$, and $D$ represent flow rate, density, kinematic viscosity, surface tension, and diffusion coefficient for the manufacturing solutions. $Re$, $Ca$, $We$, and $Pe$ are the dimensionless Reynolds, capillary, Weber, and Peclet

numbers calculated from the basic fluid properties and flow characteristics. Finally, $d_I$ and $d_O$ are the fiber exterior and interior diameters while $\Phi_f$ is the fiber porosity.

Fig. 8 presents the architecture for DNNs used to generate design models (i.e., Cases V and VI).

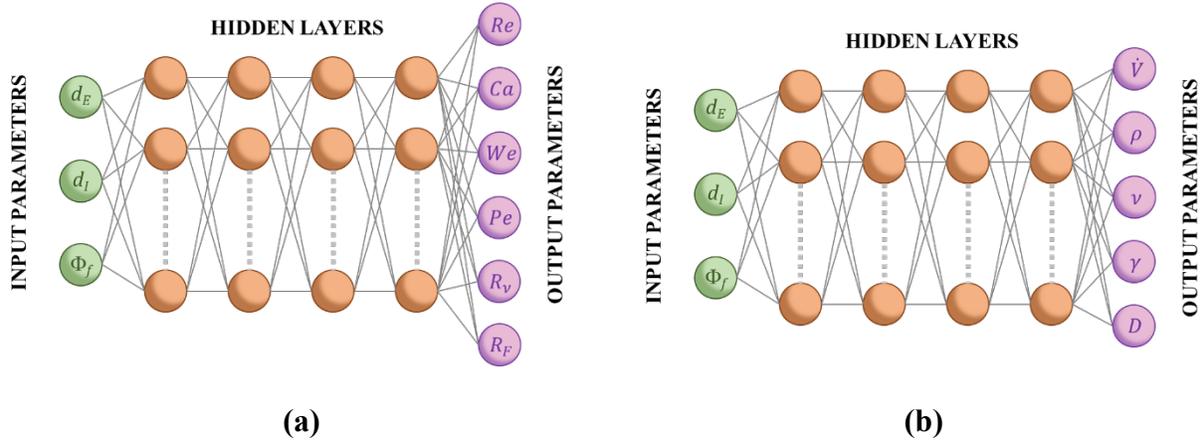

**Fig. 8.** Deep neural network architectures used to generate design models for selecting solution flow rates and fluid properties based on desired fiber characteristics of inner/outer diameter and porosity. (a) Case V: Generation of model to determine solution Reynolds, capillary, Weber, and Peclet numbers along with sheath/prepolymer viscosity and flow ratios. (b) Case VI: Generation of model to determine solution flow rates and fluid properties.

### 2.2.4. Model Performance

The DNN-based models are assessed for their performance in predicting a particular output parameter using the following definition of error, where $P_P$ is the predicted output feature value (internal/external cross-sectional diameter, porosity) and $P_T$ is the test value:

$$Error = \frac{P_P - P_T}{P_T} \times 100\% \quad (14)$$

This definition of error is applied to all computed features, whether part of a predictive model or one used to design microfiber manufacturing parameters. Note that for purposes of this study, $P_P$ and $P_T$ are average values for a given set of parameters (refer to Table 1).

3. **Results and Discussion**

All predictive and design models were generated using several datasets of varying sizes. This was done in order to study the accuracy for each level of data availability. Specifically, the following dataset sizes were investigated: 250, 500, 1,000, 2,000, 5,400, and 54,000. Additionally, different epoch sizes were used in order to aid in DNN convergence. Table 8 defines relevant parameters for each dataset used for training and validating predictive models.

**Table 8.** Parameters for datasets used to generate predictive models.

| Dataset | Number of Datapoints | Epochs |
|---|---|---|
| 1 | 250 | 8 |
| 2 | 250 | 16 |
| 3 | 500 | 8 |
| 4 | 500 | 16 |
| 5 | 1,000 | 8 |
| 6 | 2,000 | 8 |
| 7 | 5,400 | 8 |
| 8 | 54,000 | 8 |

Early testing of the DNN used to generate design models showed nonconvergence of a solution for the same number of epochs used for predictive model generation. These were therefore increased to the values shown in Table 9.

**Table 9.** Parameters for datasets used to generate Design models.

| Dataset | Number of Datapoints | Epochs |
|---|---|---|
| 1 | 250 | 16 |
| 2 | 250 | 32 |
| 3 | 500 | 16 |
| 4 | 500 | 32 |
| 5 | 1,000 | 16 |
| 6 | 1,000 | 32 |
| 7 | 2,000 | 16 |
| 8 | 2,000 | 32 |
| 9 | 5,400 | 16 |
| 10 | 5,400 | 32 |
| 11 | 54,000 | 16 |

The remainder of this section presents and discusses the results for each condition.

## 3.1. Predictive Model

Percentage error in the prediction of porosity for each of the nine datasets introduced in Table 1 is shown in Fig. 9. Results were produced from the four predictive model architectures developed in Section 2.2.3.2 and depicted in Fig. 7.

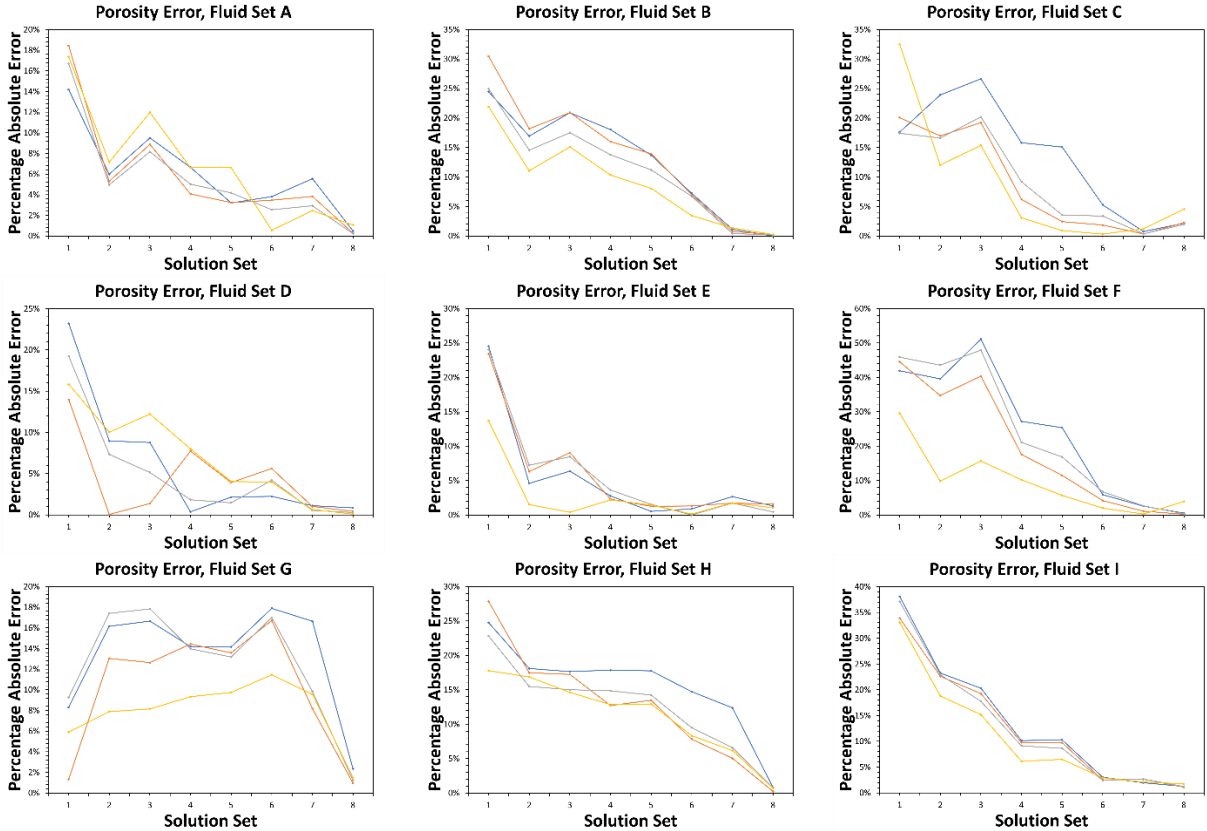

**Fig. 9.** Comparison of predictive error for porosity for all datasets defined in Table 8.

Results indicate that the predictive model increases in accuracy when it has access to a larger number of input datapoints, as would be expected. Also evident in the plots is that the inclusion of dimensionless parameters tends to increase accuracy, particularly for smaller dataset sizes, as seen for fluid sets B, C, E, F, G, H, and I. With these, training/validation of the model with all physical and derived parameters at the input and derived parameters injected additionally into Layers 2 and 3 appears to be most accurate overall. However, this is not strictly true, as can be seen with fluid sets A and D, where the inclusion of dimensionless numbers actually decreases model accuracy for smaller training/validation dataset sizes.

Percentage error for fiber inner diameter predictions for each of the nine datasets is provided in Fig. 10.

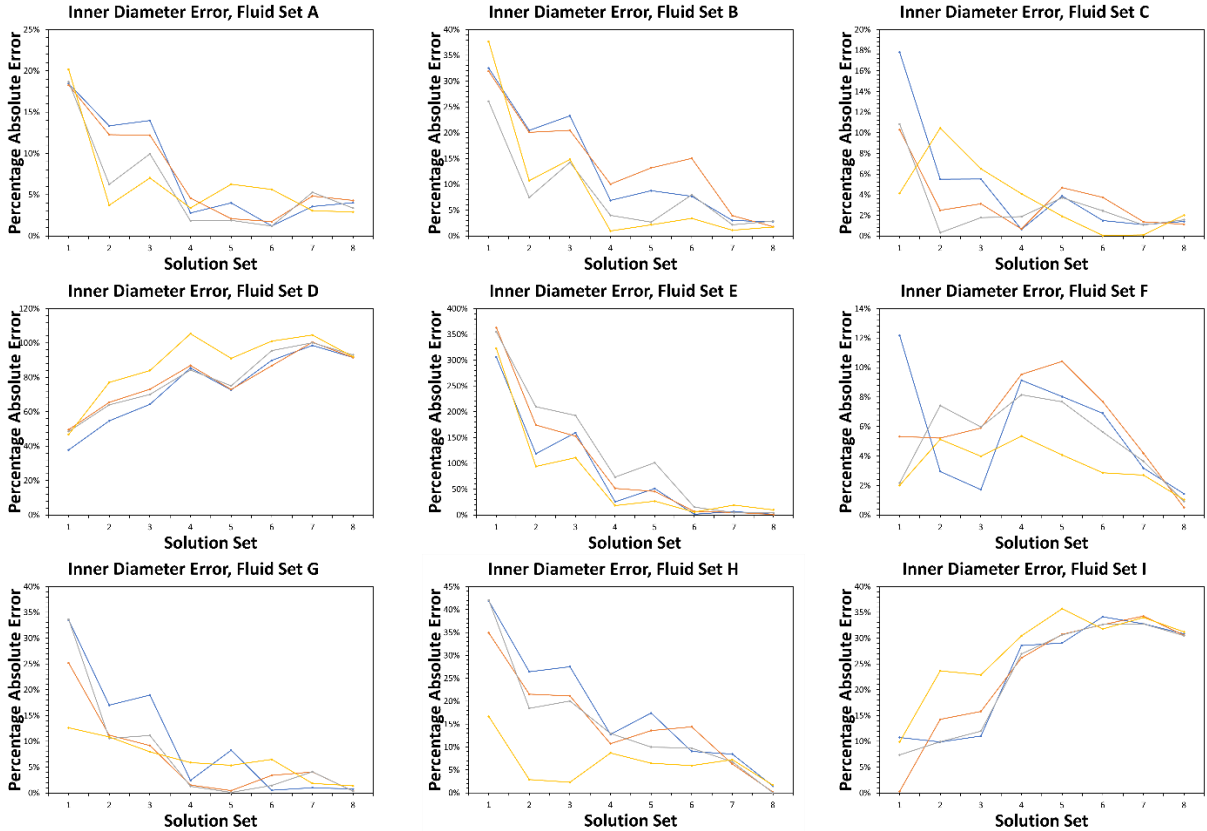

→I. Raw Input Parameters →II. Dimensionless Parameters →III. Combined Parameters →IV. Multi-Layer Physics
**Fig. 10.** Hollow microfiber inner diameter prediction error for all datasets defined in Table 8.

For this parameter, while overall the inclusion of dimensionless numbers in the training/validation datasets again appears generally to increase accuracy for smaller dataset sizes, the effect neither as strong nor as consistent. In fact, for datasets D and I, the consequence of adding physics into the dataset is nearly negligible, except that in these cases where dimensionless parameters are injected into intermediate DNN layers the results are actually notably worse overall.

Finally, the predictive results for microfiber outer diameter are provided in Fig. 11.

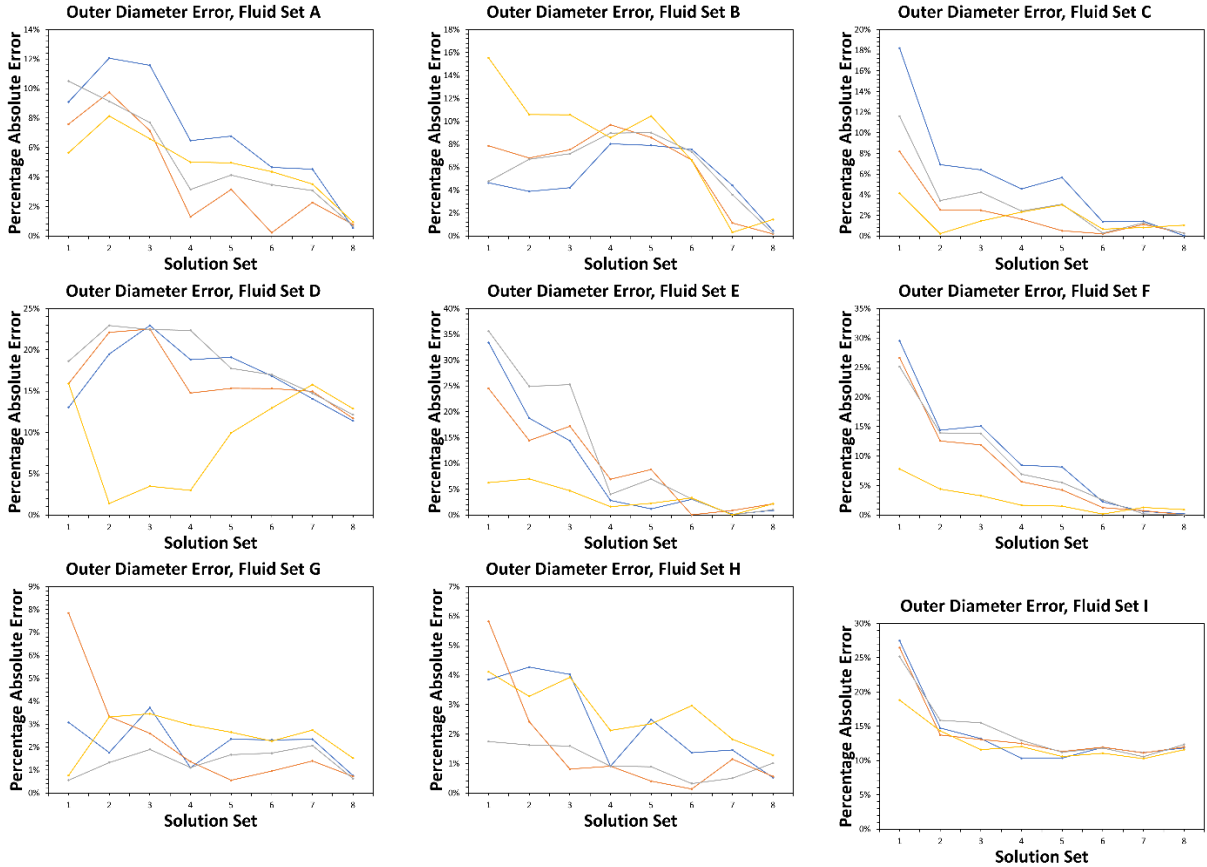

→ I. Raw Input Parameters → II. Dimensionless Parameters → III. Combined Parameters → IV. Multi-Layer Physics

**Fig. 11.** Hollow microfiber outer diameter prediction error for all datasets defined in Table 8.

As with the interior diameter, this feature sees improvement with the addition of physics for many of the fluid sets, yet again with mixed results. For example, while all sets except B appear to benefit by the addition of dimensionless numbers for small training/validation dataset sizes, the effect is nearly negligible for I. Also, for sets G and H, the combining of parameters only at the input to the DNN seems to positively impact the predictive power of the resulting model. For these sets, injecting physics into intermediate DNN layers causes the resulting model to have reduced accuracy.

## 3.2. Design Model

Results for model selection of solution Reynolds number, capillary number, Weber number, Peclet number, and sheath/prepolymer viscosity and flow rate ratios are provided in Fig. 12 through Fig. 15.

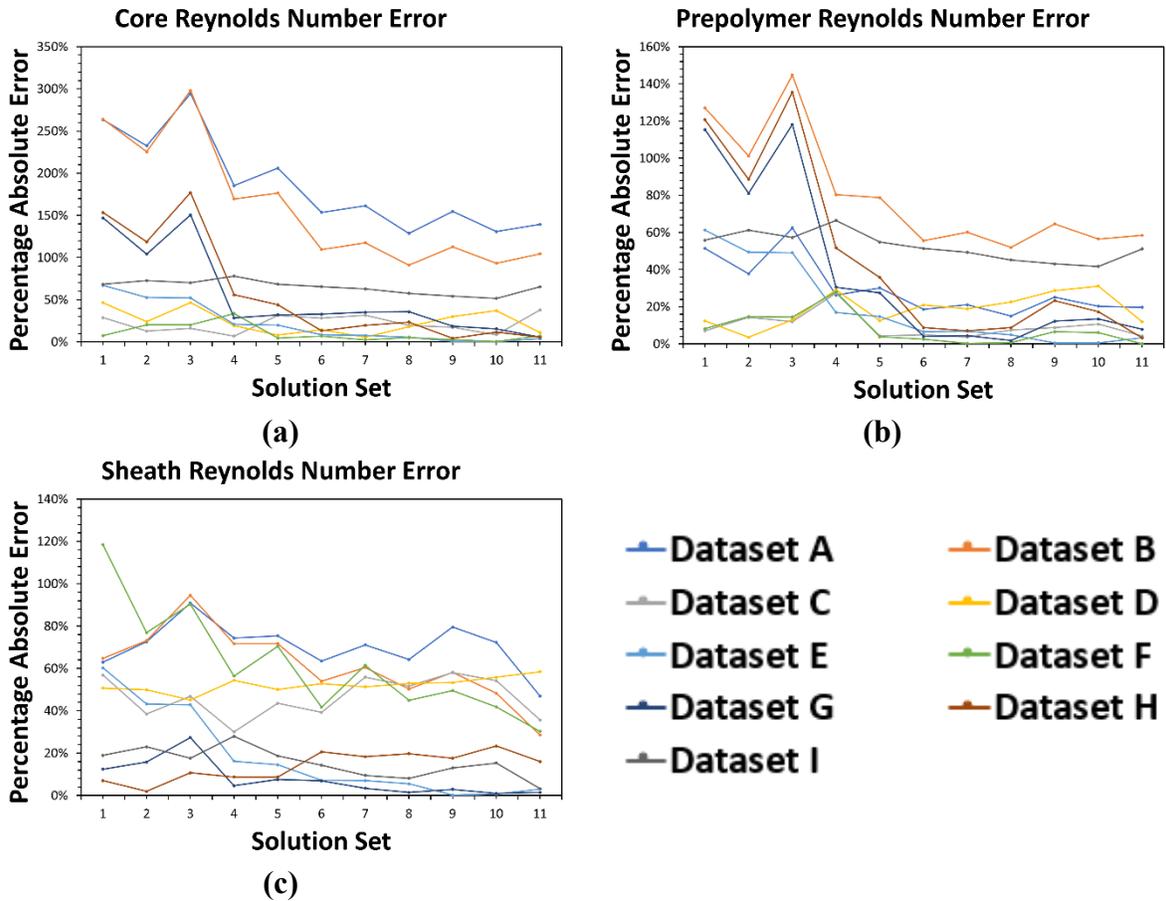

**Fig. 12.** Case V Reynolds number selection using design models developed for all nine datasets: (a) core solution, (b) prepolymer solution, and (c) sheath solution. Solution sets 1 through 11 correspond to the definitions provided in Table 9.

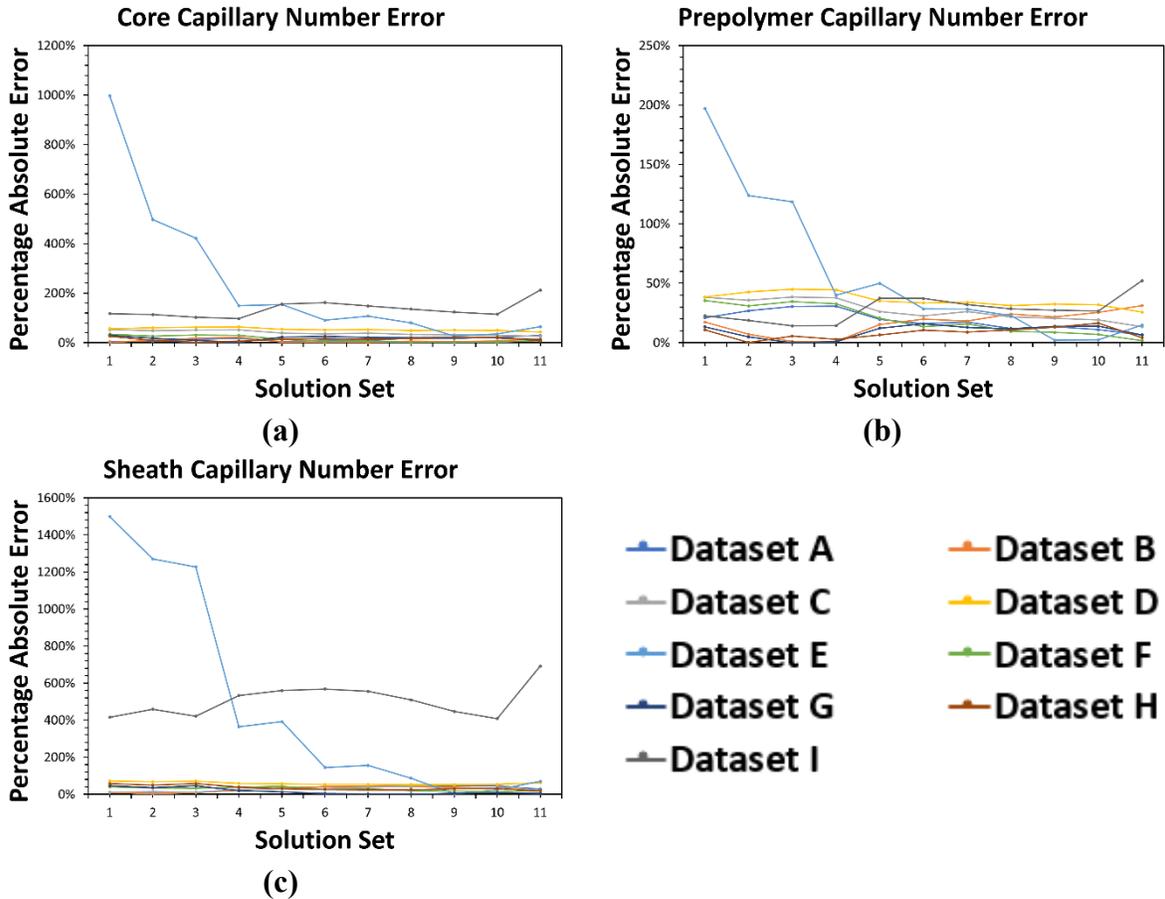

**Fig. 13.** Capillary number selection using design models developed for all nine datasets: (a) core solution, (b) prepolymer solution, and (c) sheath solution. Solution sets 1 through 11 correspond to the definitions provided in Table 9.

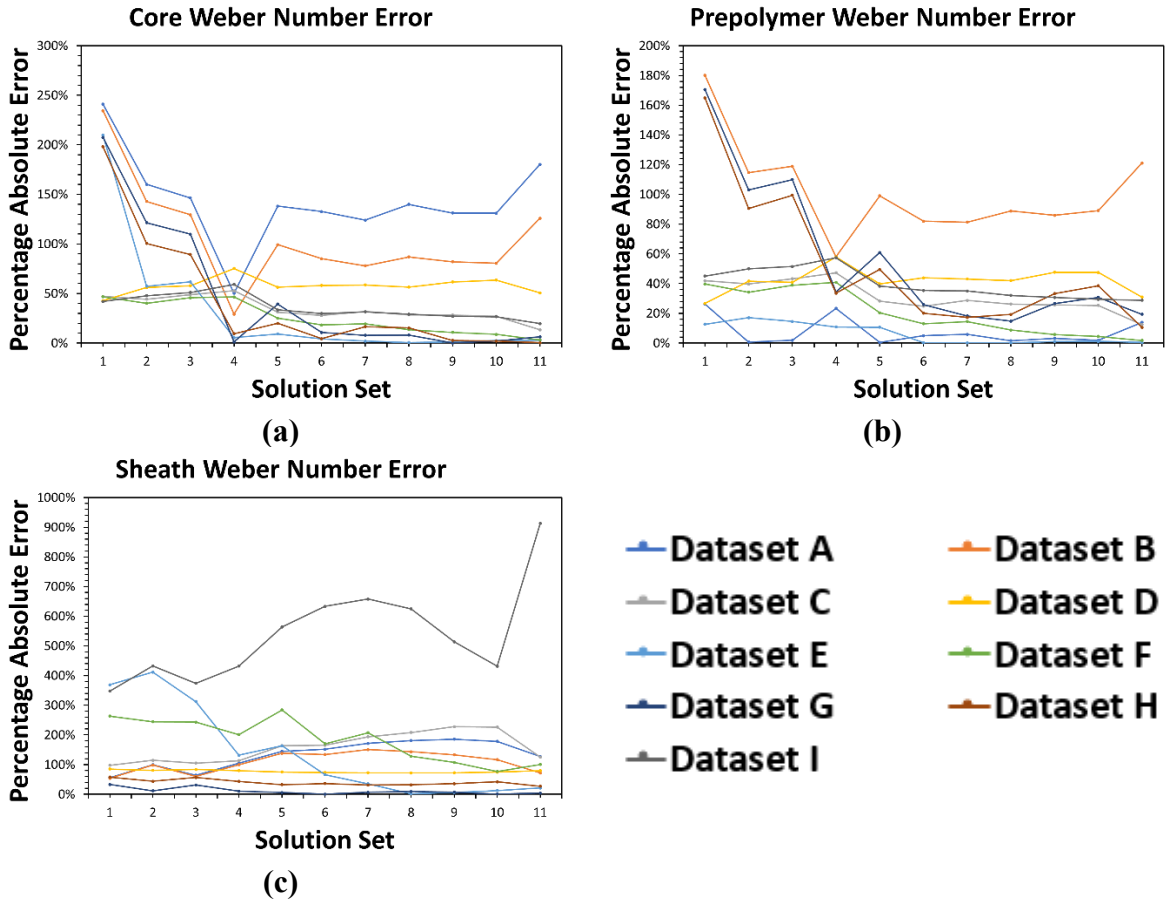

**Fig. 14.** Weber number selection using design models developed for all nine datasets: (a) core solution, (b) prepolymer solution, and (c) sheath solution. Solution sets 1 through 11 correspond to the definitions provided in Table 9.

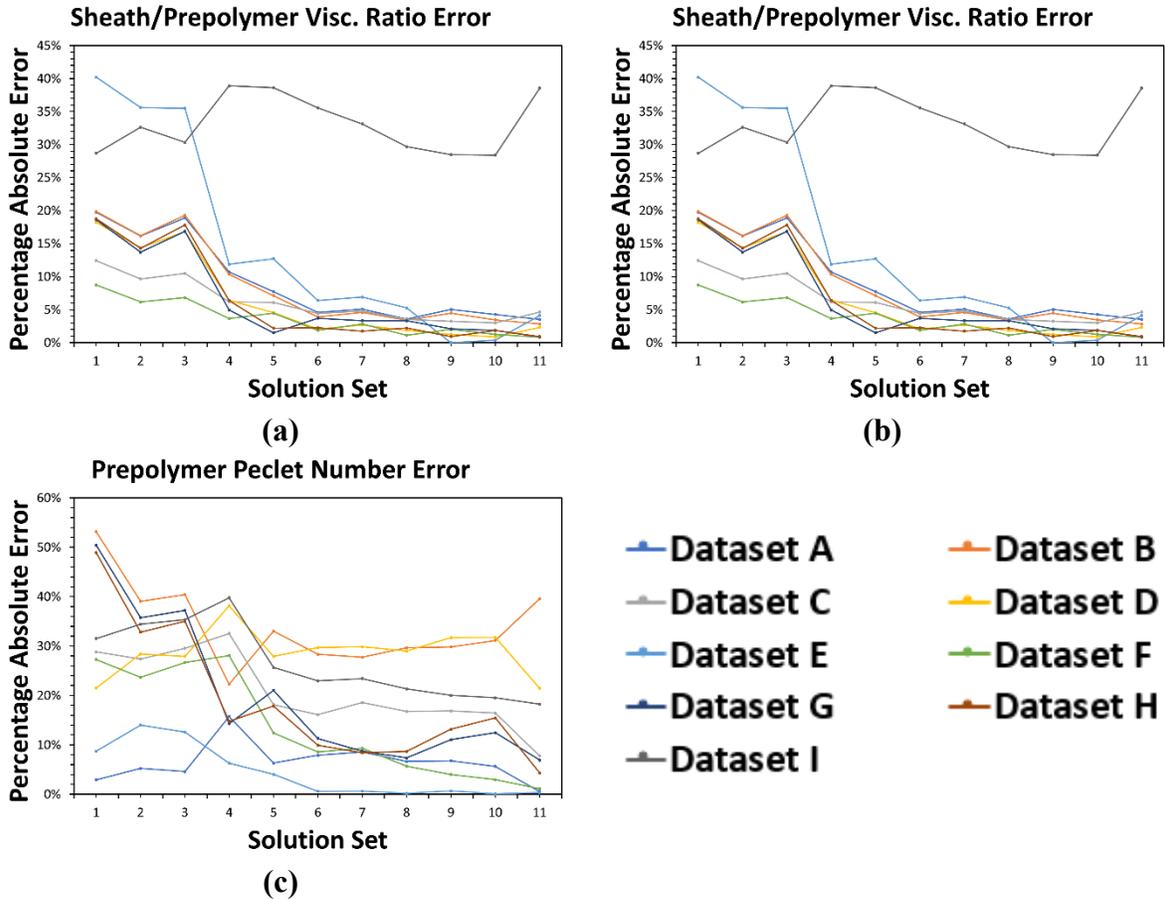

**Fig. 15.** Parameter selection for using design models developed for all nine datasets: (a) sheath/prepolymer kinematic viscosity ratio, (b) sheath/prepolymer flow rate ratio, and (c) prepolymer Peclet number. Solution sets 1 through 11 correspond to the definitions provided in Table 9.

Note the relatively poor performance of the model for selecting any of these numbers that would lead desired fiber performance. An important observation from these results is that very little improvement is even made for larger training/validation dataset sizes, signifying that the DNN is incapable to draw concrete correlations among inputs and outputs.

Results for model selection of solution flow rates, density, kinematic viscosity, surface tension, and diffusion coefficient based on desired fiber porosity and inner/outer diameters are provided in Fig. 16 through Fig. 19.

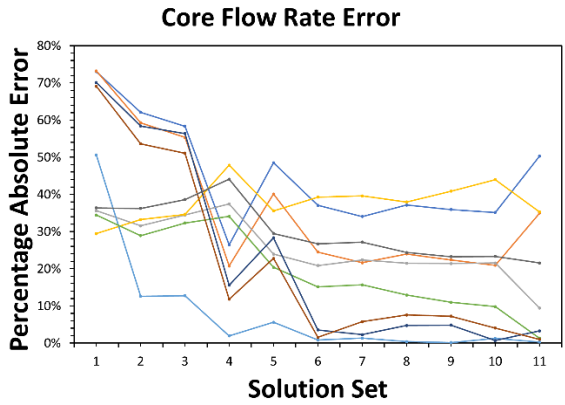
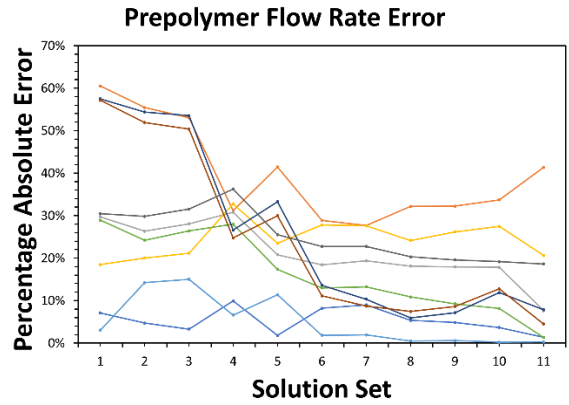
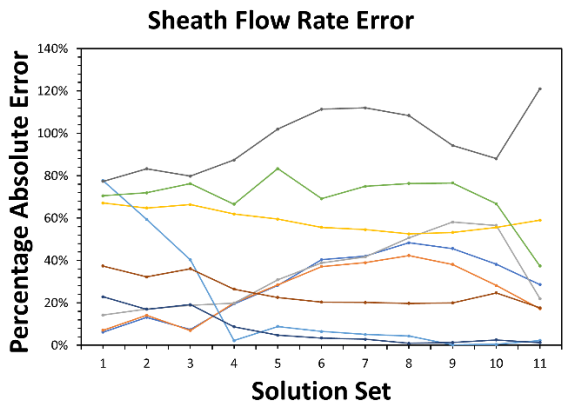
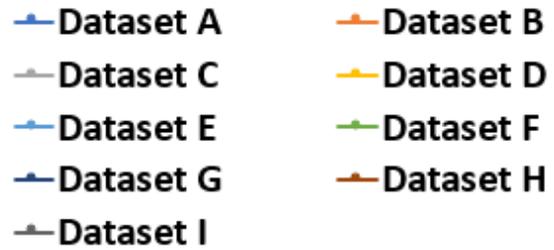

**Fig. 16.** Case VI solution flow rate selection using design models developed for all nine datasets: (a) core solution, (b) prepolymer solution, and (c) sheath solution. Solution sets 1 through 11 correspond to the definitions provided in Table 9.

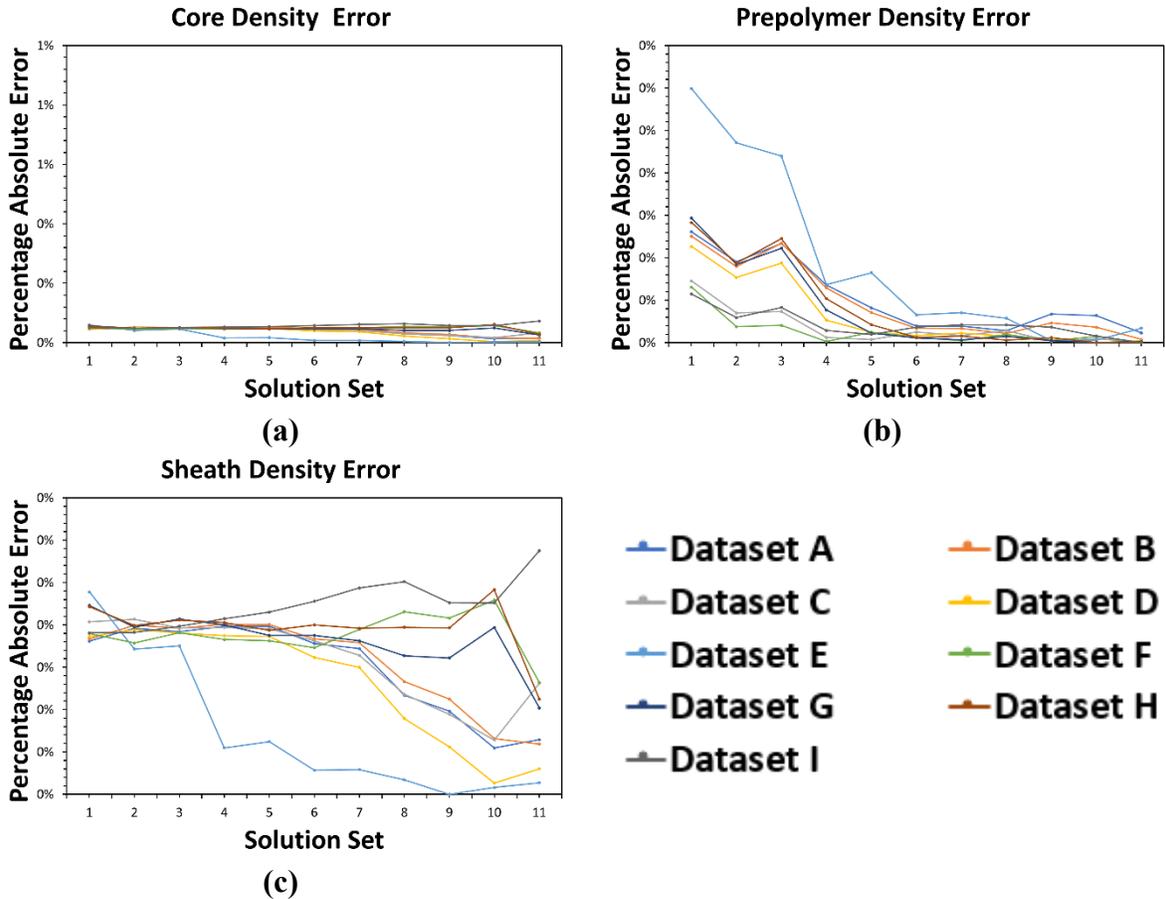

**Fig. 17.** Case VI solution density using design models developed for all nine datasets: (a) core solution, (b) prepolymer solution, and (c) sheath solution. Solution sets 1 through 11 correspond to the definitions provided in Table 9.

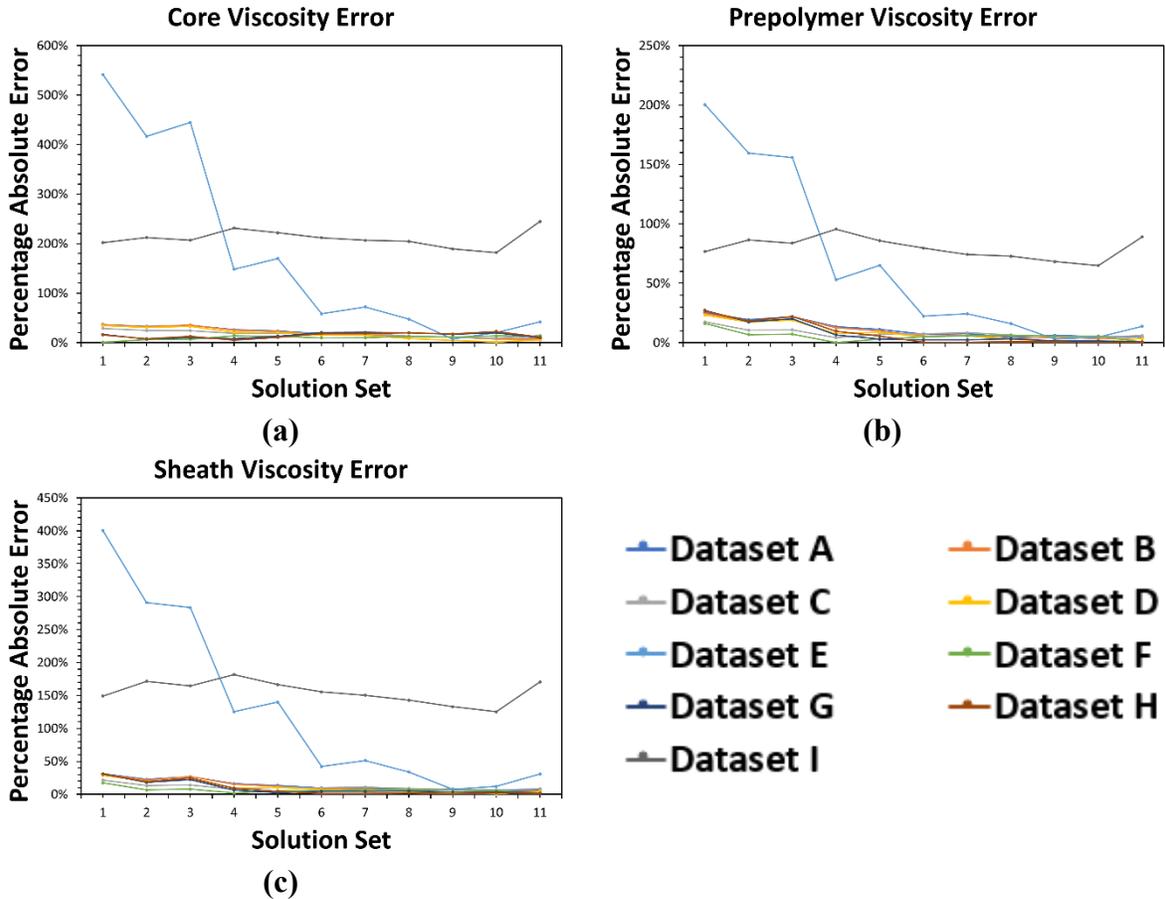

**Fig. 18.** Case VI kinematic viscosity selection using design models developed for all nine datasets: (a) core solution, (b) prepolymer solution, and (c) sheath solution. Solution sets 1 through 11 correspond to the definitions provided in Table 9.

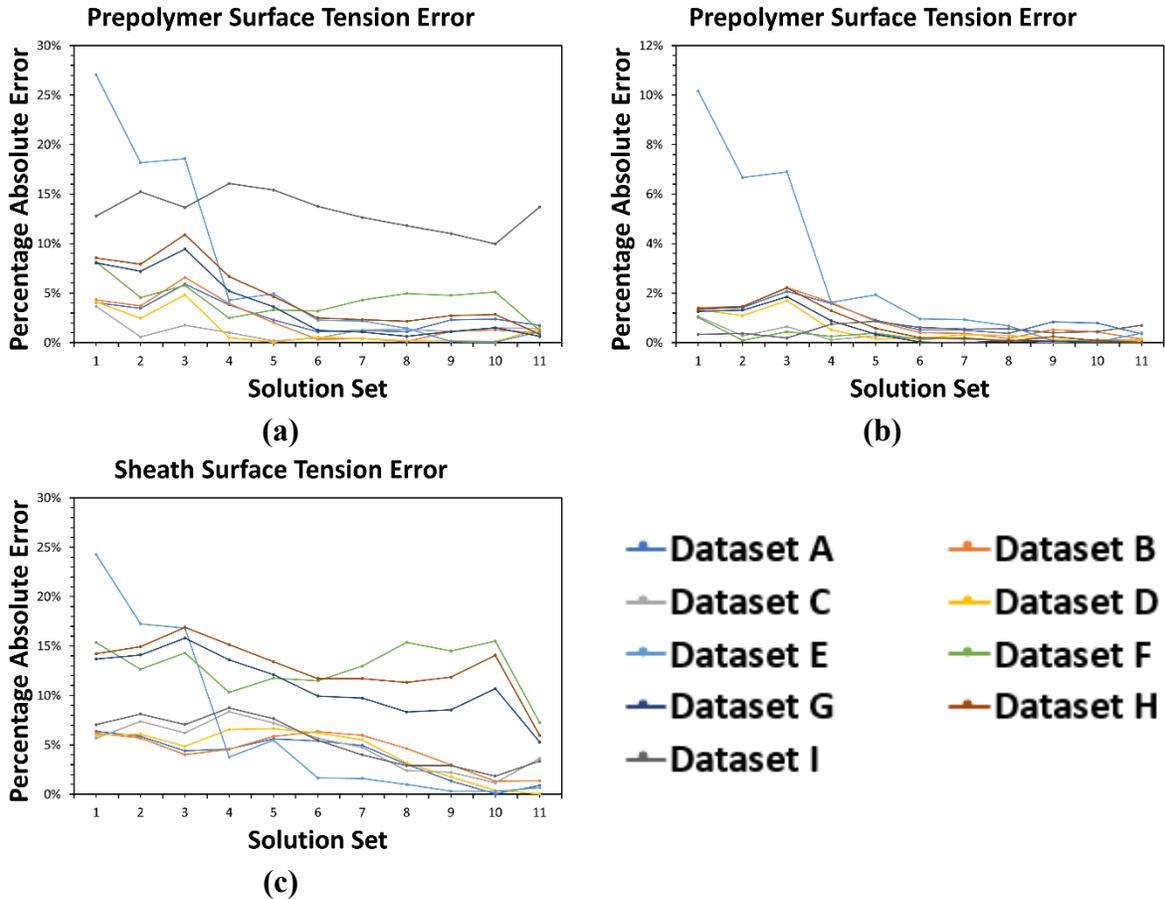

**Fig. 19.** Case VI surface tension selection using design models developed for all nine datasets: (a) sheath/prepolymer kinematic viscosity ratio, (b) sheath/prepolymer flow rate ratio, and (c) prepolymer Peclet number. Solution sets 1 through 11 correspond to the definitions provided in Table 9.

As with the selection of physically-defined parameters, selection of flow rates based on desired fiber porosity and inner/outer diameters is poor using the DNN-developed models. Interestingly, fluid properties—particularly density and surface tension—are much more accurately predicted. However, this is unsurprising based on the fact that these values show little variation across solution types and concentrations, as evidenced by a review of Table 2. In fact, viscosity selection appears to have low error except for those fluids with significantly small values, i.e., those used with a 2% and 2.5% alginate solution. It is important to note here that these two conditions are outliers of sorts—of the nine manufacturing parameters sets investigated only two do not use 3.5% alginate as the basis for generation of fibers. This may indicate that a larger number of sets with differing solution concentrations is required for adequately training and validating a design DNN-based model.

## 4. Conclusions

While not strictly true for every parameter investigated, introduction of physically-relevant relationships—specifically the Reynolds, capillary, Weber, and Peclet numbers, along with sheath/prepolymer viscosity and flow rate ratios—appears to have an overall positive impact to the accuracy of predictive models developed through a deep neural network schema. Although the general trend of increased predictive accuracy with increased dataset size is evident (as expected), for the majority of cases it appears that the addition of parameters rooted in physical relationships among manufacturing variables provides some level enhancement to a DNN-based predictive model. It is expected that accuracy will further improve as the variety and number of samples used to train and validate a DNN-based model increases.

Unfortunately, the present study did not see improvement for DNN-based models developed to select manufacturing parameters based on desired fiber characteristics. This may be due to one or more issues. While the models were accurate in selecting density, surface tension, and kinematic viscosity to some extent, it is likely due to the fact that this is due to the very low variability among these parameters. Also, as noted during the discussion, the two datasets for which viscosity was least accurately determined were for the 2% and 2.5% alginate cases, which were outliers of sorts. This may indicate that more samples comprising these fluids are necessary for training/validating a DNN that will give improved accuracy.

While more data for the diversity of datasets used to train a DNN is important to pursue for future development of a model that can select fiber manufacturing parameters, some other improvements that can be made in the process manufacture, data collection, and analysis. These are briefly discussed.

*Image-Based Porosity Evaluation.* The methodology for determining porosity that was described in Section 2.1.4 has significant potential for allowing microfibers previously fabricated and evaluated under a scanning electron microscope to be further scrutinized to determine porosity. However, the process by which this was accomplished resulted in some nonporous regions being identified by the program as being porous due to the reliance on thresholding and image contrast. This resulted in significant manual updating of the images to mask nonporous regions in order to achieve more accurate porosity values. Future work should focus on removing such limitations to reduce manual postprocessing while enhancing accuracy.

*Data Expansion*. One limitation with the data expansion methodology that was previously noted is that parameters are largely treated independently from one another. Specifically, the approach used does not correlate cross-sectional dimensions for a particular fiber with its porosity, thereby treating them as completely independent parameters. This may result in dimension-porosity combinations that are outside what might be observed for a single fiber. This potential limitation on the data expansion method should be further studied to ensure a proper bounding on the synthesized datasets.

**Acknowledgments**

This work was partially supported by the Air Force Office of Scientific Research grant FA95502410298 and the National Science Foundation, Directorate for Technology, Innovation and Partnerships grants 2014346 and 2321975. The authors would like to thank Grace Panek and Merlick Mongbo for their assistance in fabrication of microfibers and generation of SEM images.

The authors also acknowledge Mychal Trznadel and Jahid Hasan for their contributions to surface tension measurements and processing raw birefringence data.